\documentclass{article}
\usepackage[utf8]{inputenc}
\usepackage{authblk}

\title{A Property Graph Type System and Data Definition Language}
\author[1]{Mingxi Wu \thanks{mingxi.wu@tigergraph.com}}
\date{October 2018}

\usepackage[numbers]{natbib}
\usepackage{graphicx}

\begin{document}

\maketitle

\begin{abstract}
We define a type system used to model property graph. This type system includes vertex type, edge type, graph type and label type. The four types are used to model the schema of a property graph. Inheritance are allowed within a type. Based on the type system, we propose the associated data definition language (DDL).
\end{abstract}
\section{Introduction}
Property graph manages data by vertices and edges\cite{pg1,pg2,pg3}. Each vertex and edge can have a property map, storing ad hoc attribute and its value. Label can be attached to vertices and edges to group them. While this schema-less methodology is very flexible for data evolvement and for managing explosive graph element, it has two shortcomings-- 1) data dependency 2) less compression. Both problems can be solved by a schema based approach. 

In this paper, a type system used to model property graph is defined. Based on the type system, the associated data definition language (DDL)   is proposed.  

\section{Property Graph}
A property graph is a directed graph consists of vertices and edges. Each vertex and edge can have $<$property, value$>$ pairs. It is a multi-graph, meaning between two vertices, there can be multiple edges. Labels can be attached to vertices and edges, and they serve as tags.

The traditional property graph is schema-less with the benefits of great flexibility– vertex or edge instances can be labeled freely, and the properties of them can be expanded or shrunk independently of each other.

The drawback of the schema-less style is data dependency. Application written based on labelling and property map can only work when the developer knows the data. For example, if data ingestion developer is independent of the application developer, then the application developer has no way to know what's ingested (label, properties) for each graph element, thus the data dependency problem. Another problem of schema-less is that it misses the opportunity of compressing more data. With a pre-defined schema, metadata and binary data are separated, and binary data can easily be compressed since each data record is structured.

We define a type system to make a property graph schema-based. 

\section{Property Graph Type System}
In TigerGraph system, we define a property graph as a collection of typed vertices and the collection of typed edges that connect the vertices. By typed, we mean there is a user pre-defined schema for each vertex/edge type. Also, an edge type can be directed or undirected. Between two vertices, there can be multiple edges, and these edges can be of the same type or different types.

Note that readers should not be confused between the type and schema object concepts. They are different. 
\begin{itemize}
   \item \emph {Type} is a set of rules that can be named and assign to a programming construct, such as an object, a chunk of binary. 
   \item \emph {Schema object} is created by DDL and resides in the catalog. Usually, the CREATE DDL create both a type and a schema object of the new type.
\end{itemize}
    
Each of the vertices and edges is of certain pre-defined type. This section describes the type system used to model a property graph. Namely, there are vertex type, edge type, graph type, and label type. Vertex types and edge types are the building block of a graph type. Label type is a semantic tag attachable to any graph elements.

\subsection{ Vertex Type}
A vertex type is used to describe the schema of a class of vertex entities. It must have
\begin{itemize}
\item a type name: an unique identifier.
\item a set of attributes (non-empty, at least one attribute serves as the primary key): each attribute has a name and its associated data type. The data type can be any ISO SQL data type, and container type (such as as Map$<$K,V$>$, List$<$T$>$, Set$<$T$>$, and Order$<$T$>$ etc., where T is the element type).
\item a primary key: one or more attributes that can uniquely identify a vertex of this type.
\item a built-in label attribute, which encodes the labels (see 3.4). 
\end{itemize}

Optionally, a vertex type can have an auto-assigned primary key
\begin{itemize}
\item UUID. see https://en.wikipedia.org/wiki/Universally\_unique\_identifier Using UUID can avoid guessing the size of a vertex type, and also has the lease leakable information.
\end{itemize}

\subsection{ Edge Type}
An edge type is used to describe the schema of a class of edge entities. It must have
\begin{itemize}
\item a type name: an unique identifier.
\item a set of attributes (possibly empty): each attribute has a name and its associated data type. The data type can be any ISO SQL data type, and container type (such as as Map$<$K,V$>$, List$<$T$>$, Set$<$T$>$, and Order$<$T$>$ etc., where T is the element type).
\item one or more pairs of source vertex type and target vertex type. E.g., an edge type can be defined by the following two (source, target) pairs:  (FROM person, TO person) and (FROM movie, TO person). 
\item a primary key: one or more attributes that can uniquely identify an edge of this type. Note that there is a default key when a pair of source and target vertices have at most one edge of the defined edge type -- the (primary key of source vertex, the primary key of the target vertex). When there are multiple edges of the defined edge type connecting a pair of vertices, additional attributes other than the combination of (the primary key of the source vertex, the primary key of the target vertex) are needed to discriminate them. 
\item direction property: directed or un-directed. If the edge type is directed, it means the edge type models an asymmetric relationship, where the edge instance always starts from a source vertex and ends at a target vertex to capture the one direction semantics. If the edge type is undirected, it models a symmetric relationship between the source and target vertex types.
\item a built-in label attribute, which encodes the labels (see 3.4). 
\end{itemize}

Optionally, an edge type may further have
\begin{itemize}
\item zero or more key: a set of attributes that uniquely identify an edge instance, regardless whether  two edges share the same pair of the source and target vertex instances. 

\item a reverse edge type: if the edge type is specified as directed, a reverse edge type shares the same schema as the forward edge type, except the direction of it is opposite to the forward edge; and reverse edge type has a unique name. This is useful when writing queries to articulate application logic, even when the engine can traverse a directed edge both ways. 
\end{itemize}

\subsection{Graph Type}
A graph type specifies a collection of vertex types and edge types that collectively depicts the graph schema. It must have
\begin{itemize}
\item a name: an unique identifier.
\item zero or more vertex type collections (such as List[] Person, List[] animal), or a pointer which can be chased to locate a vertex type collection.  
\item zero or more edge type collections, or a pointer which can be chased to locate an edge type collection: if an edge type collection is included in a graph type, its associated source and target vertex type collections must be included in the graph.
\item zero or more graph types, or a pointer which can be chased to locate an graph type instance. 
\item a built-in label attribute, which encodes the labels (see 3.4). 
\end{itemize}

Note when a graph type is created with just a unique name, without any types in it, it is an empty graph type. In practice, user can use this type of graph to  reserves a graph name space which is global unique.  After that, user can keep adding vertex/edge types into this graph instance, thus, this graph instance's type is changed as data grow.  

\subsection{ Label Type}
Labels are semantic tags to graph data elements or a set of attributes. It has
\begin{itemize}
    \item a name: an unique identifier
    \item a description (optional): a string to describe its meaning, could be empty.
    \item a set of attributes (optional): it can be associated with a group of attributes, where each attribute has a  name and a type. Within the group, the attribute name is unique. 
\end{itemize}
This is useful when user want to persist clustering or classification result without changing the vertex, edge or graph types. 

\subsection{Type Inheritance}
For each of the above type, we allow type inheritance. Each type can have subtypes. Each subtype can only have one super type (except label type).

\subsubsection{Vertex Type Inheritance}
A vertex type can be derived from another vertex type. For example, a professor is a sub vertex type, extended from the person super vertex type. A sub vertex type can only inherit one super vertex type. The sub vertex type will
\begin{itemize}
    \item Inherit all of the super vertex type attributes 
    \item Share the same primary key of the super type. This is important to support polymorphism. In graph database, a unique identifier is important to identify an object of a particular type. To support using super type to do polymorphism, a shared primary key between the super type and its descendant types is a must.
    \item When the super type attributes are changed in the super type, all descendant types' attributes are changed accordingly.
    \item The super type attributes cannot be changed directly in its descendant subtype.
\end{itemize}

\subsubsection{Edge Type Inheritance}
An edge type can also have sub types. The sub type will
\begin{itemize}
\item Inherit all the super edge type attributes.
\item Inherit the pairs of source vertex type and target vertex type.
\item Inherit the direction property.
\item Inherit the discriminator if there is one.
\item Create the reverse edge type with a different name.
\item When the super type attributes are changed in the super type, all descendant types' attributes are changed accordingly.
\item When the super type's source and target vertex type pairs are changed in the super type, all descendant types' are changed accordingly.
\item The super type attributes and (source, target) vertex pairs cannot be changed directly in the subtype.
\end{itemize}

\subsubsection{Graph Type Inheritance}
A graph can also have sub types. The subtype will
\begin{itemize}
\item Inherit all the vertex types, edge types and graph types in the super graph type.
\item When any types contained in the super graph type change, the subgraph type changes accordingly.
\item The super graph's vertex, edge and graph types cannot be changed directly in the sub graph type.
\end{itemize}

\subsubsection{Label Type Inheritance}
A label can also have sub types. The subtype will form an inheritance relationship with the super type. A label sub type can inherit multiple super types.

\subsection{Create Statement}
A CREATE statement will create a schema object, and a schema type. The schema object will be stored in the database's catalog; its corresponding type is also created implicitly. 

\begin{itemize}
    \item create vertex 
\begin{verbatim}
#a person schema object is created; its type is person vertex type.
CREATE VERTEX person (name STRING NOT NULL PRIMARY KEY, age INT, 
gender STRING, state STRING)

#another way to specify primary key
CREATE VERTEX person (first_name STRING NOT NULL, 
last_name STRING NOT NULL, age INT, gender STRING, state STRING, 
PRIMARY KEY(first_name, last_name))

#a professor schema object is created; its type is professor vertex type
#which is a subtype of person vertex type.
CREATE VERTEX professor EXTENDS person (position STRING)
\end{verbatim}

\item create edge
\begin{verbatim}
#an friendship schema object is created; its type is friendship edge type.
CREATE UNDIRECTED EDGE friendship ((FROM person, TO person), 
(FROM person, To animal), (FROM animal, TO animal), connect_day DATETIME)

#supervise and supervised_by are both created; their type is 
#supervise and supervised_by edge type, respectively.
CREATE DIRECTED EDGE supervise (FROM person, TO person, 
connect_day DATETIME) WITH REVERSE_EDGE=”supervised_by”

CREATE DIRECTED EDGE supervise (FROM person, TO person,
connect_day DATETIME, DISCRIMINATOR (connect_day))
 WITH REVERSE_EDGE=”supervised_by”

#a mentorship schema object is created; its type is 
mentorship edge type, which is a subtype of supervise
# edge type.
CREATE DIRECTED EDGE mentorship EXTENDS supervise(end_day DATETIME)
WITH REVERSE_EDGE= ”mentored_by”
\end{verbatim}
\item create graph
\begin{verbatim}
#below two graph schema objects are created. Both contain 
#the same schema object of person vertex type.
CREATE GRAPH social (person, friendship)
CREATE GRAPH company (person, supervise)

#below, a graph facebook is created based on social, 
#it has alumni edge type added.
CREATE UNDIRECTED EDGE alumni_relation (FROM person, TO person)
CREATE GRAPH facebook EXTENDS social (alumni_relation)
\end{verbatim}
\item create label
\begin{verbatim}
CREATE LABEL color Description "color super class”
CREATE LABEL car Description "car super class”
CREATE LABEL redcar EXTENDS color, car 
\end{verbatim}
\end{itemize}

\subsection{Drop Statement}
A DROP statement will remove a schema object, and its corresponding type from the catalog.

For inheritance constrain related dropping: 
\begin{itemize}
\item[-]When dropping a supertype A, an error is raised if there exists subtypes B of A. 
\item[-]When dropping a subtype B, all data of supertype A stays.
\end{itemize}

The common drop statements.

\begin{itemize}
\item drop vertex
\begin{verbatim}
#Drop vertex type  can use CASCADE option. 
#When dropping a vertex type V, without CASCADE keyword,
#an error will be raised if there is an edge type  E 
#referencing this vertex type V. With CASCADE keyword,  
#E will be modified to reflect the disappearance of V.  
#When the source types  (or the target types) of E mention only 
#V, then E is dropped.
DROP VERTEX person CASCADE
DROP VERTEX person, city, school

#delete all vertex schema objects and their types
DROP VERTEX *
\end{verbatim}
\item drop edge
\begin{verbatim}
#drop the edge type and object
DROP EDGE friendship, supervise
#delete all edge types
DROP EDGE *
\end{verbatim}

\item drop graph

\begin{verbatim}
DROP GRAPH social, company 
\end{verbatim}

\item drop label

\begin{verbatim}
DROP LABEL red
DROP LABEL color
\end{verbatim}
\end{itemize}

\subsection{Alter Statement}
An ALTER statement will change a schema object and its type.
\begin{itemize}
\item alter vertex add/drop attributes
\begin{verbatim}
ALTER VERTEX person ADD (ssn VARCHAR(9))
ALTER VERTEX person DROP (ssn VARCHAR(9))
\end{verbatim}
\item alter edge add/drop attributes
\begin{verbatim}
ALTER EDGE friendship ADD (location VARCHAR(20))
ALTER EDGE friendship DROP (location VARCHAR(20))
\end{verbatim}
\item  alter graph add/drop vertex type and edge type
\begin{verbatim}
ALTER GRAPH school ADD VERTEX (professor, student)
ALTER GRAPH school DROP VERTEX (professor)

#note below, it implicitly add all dependent source vertex
#type and target vertex type into the graph
ALTER GRAPH school ADD EDGE (teach_class)

#only drop edge type
ALTER GRAPH school DROP EDGE (teach_class)

#note below, it will drop all dependent edges from the graph.
ALTER GRAPH school DROP VERTEX (professor) CASCADE
\end{verbatim}
\end{itemize}

\section{Multiple Graph Instances Support}
In practice, user can pre-define all the needed types, and then create  instances of graph. 

\noindent For example, we first define a set of types. 
\begin{itemize}
\item a  vertex type A
\item a  edge type B
\item a graph type G containing a A vertex type collection and a B edge type collection.
\end{itemize}

Next, when user defines an instances for graph type G, this instance will have a storage container containing the storage container of the vertex instances of type A, and the storage container of the edge instances of type B. Graph element(vertices and edges) can be inserted into container A and container B. 

However, most database DDLs accomplish the above two steps in a combined single step. There, the graph schema object represents the graph instance, the vertex/edge schema object represents the container of the vertex/edge instances. DDL usually has the format "CREATE XXX name", which will create a schema object with the specified name residing in the database catalog. 
Each schema object corresponds to a storage container in the database storage layer. For example, given the following set of DDL commands

\begin{itemize}
\item CREATE VERTEX A(...)
\item CREATE UNDIRECTED EDGE B(FROM A, TO A)
\item CREATE GRAPH G1(references A, references B)
\end{itemize}

\noindent After executing them, the database catalog will have 
\begin{itemize}
\item a  vertex type A
\item a schema object A of vertex type A
\item a  edge type B
\item a schema object B of edge type B
\item a graph type G1
\item a schema object G1 (of graph type G1), which contains a reference G1.A pointing to the schema object A, and a reference G2.B pointing to the schema object B
\end{itemize}

The vertex (edge) schema object will be associated with a storage container storing the collection of the  vertex (edge) type instances. The graph schema object will be associated with a storage container which 
contains the references to the vertex and the edge type storage containers.

What if we want to create a graph that has its own vertex type A and edge type B containers? With the following DDL, we can accomplish it. 

\begin{itemize}
\item CREATE GRAPH G2(A, B)
\end{itemize}

\noindent After executing it, the database catalog will have 
\begin{itemize}
\item a  vertex type A
\item a schema object A of vertex type A
\item a  edge type B
\item a schema object B of vertex type B
\item a graph type G1
\item a schema object G1 (of graph type G1), which contains a reference G1.A pointing to the schema object A, and a reference G2.B pointing to the schema object B
\item a graph type G2
\item a schema object G2.A of vertex A
\item a schema object G2.B of edge type B
\item a schema object G2 of graph type G2
\end{itemize}

\noindent Can we create a graph of an known type? We can. By exuecting the following command

\begin{itemize}
\item CREATE GRAPH G3 as G1
\end{itemize}

An instance of graph type G1 is created. The new instance is named G3. At this point, the catalog has 

\begin{itemize}
\item a  vertex type A
\item a schema object A of vertex type A
\item a  edge type B
\item a schema object B of vertex type B
\item a graph type G1
\item a schema object G1 (of graph type G1), which contains a reference G1.A pointing to the schema object A, and a reference G2.B pointing to the schema object B
\item a graph type G2
\item a schema object G2.A of vertex A
\item a schema object G2.B of edge type B
\item a schema object G2 of graph type G2
\item a schema object G3 (of graph type G1), which contains a reference G3.A pointing to the schema object A, and a reference G3.B pointing to the schema object B
\end{itemize}

\noindent Further, if a user invoke another DDL command to create graph G4, which references G1 type as below 

\begin{itemize}
\item CREATE GRAPH G4(references G1)
\end{itemize}

\noindent After executing it, the database catalog will have 
\begin{itemize}
\item a  vertex type A
\item a schema object A of vertex type A
\item a  edge type B
\item a schema object B of vertex type B
\item a graph type G1
\item a schema object G1 (of graph type G1), which contains a reference G1.A pointing to the schema object A, and a reference G2.B pointing to the schema object B
\item a graph type G2
\item a schema object G2.A of vertex A
\item a schema object G2.B of edge type B
\item a schema object G2 of graph type G2
\item a schema object G3 (of graph type G1), which contains a reference G3.A pointing to the schema object A, and a reference G3.B pointing to the schema object B
\item a graph type G4
\item a schema object G4 of graph type G4, which contains a reference G4.G1 pointing to the schema object G1
\end{itemize}

Note the above, schema object must have a unique global name. One convention is to chain an object's ancestor's schema object name by ".". 

At data manipulation time, user can use the following syntax to access a graph data.
\begin{itemize}
\item Use GRAPH name
\end{itemize}

Inside a graph, user can continue to change the graph schema and create vertex types. 

\begin{itemize}
\item Use GRAPH G1
\item CREATE VERTEX C (...)
\end{itemize}

At this point, the catalog will have 
\begin{itemize}
\item a  vertex type A
\item a schema object A of vertex type A
\item a  edge type B
\item a schema object B of vertex type B
\item a graph type G1
\item a vertex type G1.C
\item a schema object G1.C of vertex type G1.C
\item a schema object G1 (of graph type G1), which contains a reference G1.A pointing to the schema object A, and a reference G2.B pointing to the schema object B, the schema object G1.C
\item a graph type G2
\item a schema object G2.A of vertex A
\item a schema object G2.B of edge type B
\item a schema object G2 of graph type G2
\item a schema object G3 (of graph type G1), which contains a reference G3.A pointing to the schema object A, and a reference G3.B pointing to the schema object B
\item a graph type G4
\item a schema object G4 of graph type G4, which contains a reference G4.G1 pointing to the schema object G1
\end{itemize}

\bibliographystyle{plain}
\bibliography{references.bib}

\section{Appendix}
\begin{verbatim}
CREATE_QB ::= ‘create’ (CREATE_VERTEX_QB | CREATE_EDGE_QB |
                        CREATE_GRAPH_QB | CREATE_LABEL_QB)

CREATE_VERTEX_QB ::= 'vertex' IDENTIFIER '(' ATTRIBUTE_ELEMENT (','       
                    ATTRIBUTE_ELEMENT)* (, ‘primary’ ‘key’ 
                    ‘(‘  IDENTIFIER (,IDENTIFIER)*’)’)?  ')' 
CREATE_EDGE_QB ::=   ('directed’ |'undirected’')  'edge' IDENTIFIER 
 '(' 'from' ((IDENTIFIER ('|' IDENTIFIER)*)|STAR) ',' 'to'
 ((IDENTIFIER ('|' IDENTIFIER)*)|STAR) (',' ATTRIBUTE_ELEMENT)* ')'  
('WITH'  KEY_VAL_SUFFIX)?

CREATE_GRAPH_QB ::= 'graph' IDENTIFIER '(' IDENTIFIER (',' IDENTIFIER)* ')'

CREATE_LABEL_QB ::= 'label' IDENTIFIER (DESCRIPTION STRING_LITERAL)? 

DROP_QB ::= ‘drop’ DROP_VERTEX_QB | DROP_EDGE_QB | DROP_GRAPH_QB | 
DROP_LABEL_QB

DROP_VERTEX_QB ::= ‘vertex’ IDENTIFIER (, IDENTIFIER)* ‘cascade’?

DROP_EDGE_QB ::=  ‘edge’ IDENTIFIER (, IDENTIFIER)* 

DROP_GRAPH_QB ::= ‘graph’ IDENTIFIER (, IDENTIFIER)* 

DROP_LABEL_QB ::= ‘label’ IDENTIFIER (, IDENTIFIER)*

ALTER_QB ::= ‘alter’ (ALTER_VERTEX_QB | ALTER_EDGE_QB|ALTER_GRAPH_QB)

ALTER_VERTEX_QB ::= 'vertex' IDENTIFIER (ALTER_ADD_VERTEX_PROP_QB | 
ALTER_UPDATE_VERTEX_PROP_QB | ALTER_DROP_VERTEX_PROP_QB)

ALTER_EDGE_QB ::= 'edge' IDENTIFIER (ALTER_ADD_EDGE_PROP_QB | 
ALTER_UPDATE_EDGE_PROP_QB | ALTER_DROP_EDGE_PROP_QB)


\end{verbatim}
\end{document}